# Linear magnetoelectric effect as a signature of long-range collinear antiferromagnetic ordering in the frustrated spinel $CoAl_2O_4$

Somnath Ghara,[1] N. V. Ter-Oganessian,[2] and A. Sundaresan[1,*]

[1]Chemistry and Physics of Materials Unit and International Centre for Materials Science, Jawaharlal Nehru Centre for Advanced Scientific Research, Jakkur P.O., Bangalore 560 064, India.

[2]Institute of Physics, Southern Federal University, Rostov-on-Don 344090, Russia.

## Abstract

The ground state of the frustrated *A*-site magnetic spinel $CoAl_2O_4$ has been a controversial issue whether it is a collinear antiferromagnetic ordering or a spiral spin - liquid state, as the ratio of the two competing interactions, $J_2/J_1$ lies close to the boundary between these two ground states. Here, we address the magnetic ground state in $CoAl_2O_4$ with different amount of $Co^{2+}/Al^{3+}$ site disorder from the study of magnetoelectric effect and Monte Carlo simulations. $CoAl_2O_4$ with low site disorder exhibits linear magnetoelectric effect below the magnetic ordering temperature. With increasing disorder, the magnetoelectric effect is suppressed and the sample with 14% disorder exhibits a spin glass behavior without the magnetoelectric effect. Monte Carlo simulations support the experimental findings and suggest that the site disorder suppresses long - range antiferromagnetic order and induces a spin glass state. Since the linear magnetoelectric effect requires a long - range magnetic ordering, we suggest that the ground state of $CoAl_2O_4$ with low site disorder is a collinear antiferromagnet.

*E-mail: sundaresan@jncasr.ac.in

## I. Introduction

Spinel compounds, $AB_2X_4$ ($X$ = O, S, Se) with magnetic ions located solely on the $A$-site have received much attention because of possible magnetic frustration arising from competing nearest-neighbor ($J_1$) and next-nearest-neighbor ($J_2$) exchange interactions. In the normal cubic spinel, $A^{2+}[B_2^{3+}]X_4$, $A^{2+}$ ions occupy tetrahedral sites and form the diamond lattice and $B^{3+}$ ions occupy octahedral sites and form the pyrochlore lattice, as shown in Fig. 1a. In the case of $A$-site magnetic spinel, the diamond lattice that is composed of two face-centered cubic (fcc) lattices with magnetic ions at (000) and (¼, ¼, ¼), provides two different exchange paths, namely, the nearest-neighbor ($J_1$) interaction, which couples ions between the two fcc lattices and next-nearest-neighbor ($J_2$) exchange interactions among ions within the same fcc lattice as shown in Fig. 1b. Thus, the magnetic interaction becomes frustrated when $J_2$ is significant. From a theoretical calculation, Bergman *et al.*, reported that, in the limit of $0 < J_2/J_1 < 1/8$, the magnetic ground state is a collinear Neel antiferromagnetic ordering (AF), while $J_2/J_1 > 1/8$ gives rise to a degenerate spiral spin - liquid state [1]. For example, $Co_3O_4$ and $MnAl_2O_4$ with $J_2/J_1 \sim 0.019$ and 0.09, respectively, exhibit a well-defined long - range collinear antiferromagnetic ordering, while $MnSc_2S_4$ has $J_2/J_1 \sim 0.85$ and exhibits a spiral spin - liquid ground state [2-6]. In addition, the anti-site disorder between $A^{2+}$ and $B^{3+}$ ions that depends on the synthetic condition also plays a major role in determining the magnetic ground state of $A$-site magnetic spinels [7]. Thus, depending on the strength of the nearest-neighbor ($J_1$) and next-nearest-neighbor ($J_2$) exchange interactions and the extent of anti-site disorder, $A$-site magnetic spinels possess versatile magnetic properties.

The $A$-site magnetic spinel, $CoAl_2O_4$ represents an interesting case where the ratio of $J_2/J_1$ ( ~ 0.109) is close to the critical ratio of 1/8, separating the collinear antiferromagnetic and the spiral

spin - liquid states [1,2]. Many contradicting experimental results have been reported regarding the true nature of the magnetic ground state of $CoAl_2O_4$. In an earlier report by W. L. Roth, using neutron diffraction experiments on a polycrystalline sample of $Co_{1-x}Al_x[Al_{2-x}Co_x]O_4$ with an anti-site disorder of $x$ = 0.05, which hereafter will be represented as CAO:0.05, a collinear antiferromagnetic ordering of $Co^{2+}$ spins at 4.2 K was suggested from the observation of a weak magnetic Bragg peak at (200) position in neutron diffraction pattern, similar to the arrangement of $Co^{2+}$ spins in $Co_3O_4$ [5]. Later, a spin glass state was proposed using magnetic, electron spin resonance and heat capacity measurements in CAO:0.08 [8]. Recently, Kenataro Hanashima *et al.* investigated a series of compounds with different degree of disorder ($x$) and have shown that the spin glass state is present with higher degree of disorder ($x \geq 0.10$), while the magnetic state changes to spiral spin - liquid state as it becomes more ordered ($x \leq 0.06$) [9]. Based on inelastic neutron diffraction experiments on CAO:0.07-0.08, Krimmel *et al.* have shown that $J_2/J_1 \sim 0.17 > 1/8$ and concluded the presence of spiral spin - liquid state with a very strong but short - range spin correlations [3]. The presence of spin - liquid like state is also supported by the results of the neutron diffraction experiments on single crystal of CAO:0.08 by Zaharko *et al.*[2]. Later, they reported that the magnetic ordering is unconventional and its bulk magnetic properties are isotropic [10]. On the other hand, it has been shown that these properties are anisotropic microscopically as inferred from neutron diffraction experiments under magnetic fields [10]. On the same sample, using local spin probe techniques such as electron spin resonance, nuclear magnetic resonance and muon spin relaxation, Iakovleva *et al.* emphasize the role of structural disorder and suggest the slowing down of spin dynamics below 100 K with a gradual crossover to a quasistatic regime below 8 K, where a short - range spin fluctuation persists [11]. In contrast, Gregory J. MacDougall *et al.*, using single crystal neutron diffraction experiments on CAO:0.02, reported that $CoAl_2O_4$ exhibits a first order magnetic phase transition; however a true long -range order is inhibited by the frozen magnetic domain walls, which cannot grow even at lowest temperature [12]. On the other hand, B. Roy *et al.* suggested a collinear antiferromagnetic ordering from the study of nuclear magnetic resonance and neutron diffraction on CAO:0.057 [13].

Recently, an emergence of linear magnetoelectric effect has been predicted theoretically in *A*-site collinear antiferromagnetic spinels and confirmed experimentally in several *A*-site magnetic spinels with different magnetic ions, $Co_3O_4$ and $MnB_2O_4$ ($B$ = Ga and Al), where the single-ion anisotropy of the magnetic ions $Co^{2+}$/$Mn^{2+}$ located at the local non-centrosymmetric crystal environment is responsible for the magnetoelectric polarization [14,15]. Since collinear antiferromagnetism is one of the essential factors to observe linear magnetoelectric effect in *A*-site magnetic spinels, we have investigated the magnetoelectric properties of $Co_{1-x}Al_x[Al_{2-x}Co_x]O_4$ ($x$ = 0.05, 0.07, 0.11 and 0.14) with different degrees of anti-site disorder to understand the nature of magnetic ground state.

Indeed, we observe linear magnetoelectric effect in $Co_{1-x}Al_x[Al_{2-x}Co_x]O_4$ with $x$ = 0.05, 0.07, and 0.11. The sample with $x$ = 0.14 exhibits a spin glass behavior without magnetoelectric effect. The observation of magnetoelectric effect, along with the dc and ac magnetization and heat capacity measurements, suggests that the magnetic ground state of the ordered CAO: $x \leq 0.11$ is a collinear antiferromagnetic state. The magnetic ground state changes from a collinear antiferromagnetic state to a spin glass state through a re-entrant spin glass state with increasing disorder. These findings are consistent with the results of Monte Carlo simulations.

## II. Experimental Details

Polycrystalline sample of $CoAl_2O_4$ was prepared by the conventional high temperature solid state reaction method. Stoichiometric amounts of $Co_3O_4$ and $Al_2O_3$ were mixed and sintered at 1130 °C in air with several intermittent grindings. The following cooling rates, 0.2 °C/min, 1 °C/min and 2 °C/min, were adapted to prepare different anti-site disordered samples. The sample with the highest disorder was prepared by a quenching process. Phase purity was confirmed using room temperature X - ray diffraction data obtained from PANalytical Empyrean alpha 1 diffractometer with a monochromatized $CuK\alpha_1$ radiation. DC magnetic measurements were carried out with a Superconducting QUantum Interference Device (SQUID-Quantum Design, USA). Temperature dependent ac susceptibility was measured with Physical Properties Measurement System (PPMS, Quantum Design - USA). Heat capacity was measured by a relaxation technique using PPMS. Temperature dependent dielectric constant and dielectric loss data at different magnetic fields were recorded with Agilent E4980A LCR meter with a heating rate of 2 K/min with the help of a multifunctional probe provided by Quantum Design in PPMS. Pyroelectric current measurements were carried out with a Keithley 6517A electrometer in PPMS with the help of the multifunctional probe. For pyroelectric current measurements, the sample was poled from 25 K to lowest temperature at various magnetic fields and an electric field of $E_P$ = + 6.5 kV/cm. After magnetoelectric poling, electric field was removed and the sample was kept shorted for enough time to remove any stray currents and the pyroelectric current was recorded while warming the sample with a rate of 8 K/min in the presence of magnetic field. In the dc-biased current measurements, the sample was cooled to the lowest temperature without magnetoelectric poling and the current was measured in the presence of an electric field of $E_{Bias}$ = + 6.5 kV/cm and a magnetic field of 80 kOe while warming with a rate of 8 K/min.

## III. Results and Discussion

Room temperature X - ray diffraction patterns, as shown in Fig. S1 in the supplemental material, confirm the phase purity of the samples [16]. The detailed structural analysis was performed with Rietveld refinement technique using FullProf software package and the crystallographic parameters obtained from the analysis are shown in Table - I in the supplemental material [16-18]. $CoAl_2O_4$ crystallizes in normal spinel structure with *Fd*-3*m* space group. As mentioned before, a tendency of cation inversion between tetrahedral and octahedral sites is observed and the degree ($x$) of the cation inversion in $Co_{1-x}Al_x[Al_{2-x}Co_x]O_4$ is controlled by the different cooling rates used in synthesis process. By varying relative occupancies of $Co^{2+}$ and $Al^{3+}$ ions in tetrahedral and octahedral sites in refinements, it is observed that the samples cooled with the rates of 0.2, 1 and 2 °C/min have a degree ($x$) of disorder of 0.05, 0.07 and 0.11 respectively, while this is 0.14 for the sample that was quenched directly to room temperature.

Temperature dependent field-cooled (FC) and zero-field-cooled (ZFC) magnetization, $M$ ($T$), measured under an applied magnetic field of 100 Oe for all the samples are shown in Fig. 2a. For $x$ = 0.05, a broad peak is observed around 14 K in both ZFC and FC data, while a sharp peak is observed at $T^*$ ~ 9 K in $dM/dT$, as shown in the inset of Fig. 2a. In earlier studies, it has been suggested that this broad peak is associated with either a spiral spin - liquid state or a long - range collinear antiferromagnetic ordering of $Co^{2+}$ spins [9,13]. Upon further cooling, a small bifurcation between ZFC and FC curve is observed. With increasing anti-site disorder ($x$), the broad peak shifts to lower temperatures and a new cusp at a further lower temperature appears for $x = 0.11$. Also the bifurcation between ZFC and FC curves increases with increasing $x$. For $x = 0.14$, the broad peak vanishes and only a sharp cusp is observed around 5 K with a strong bifurcation between FC and ZFC curves. The ac susceptibility measurements indicated that this cusp is associated with spin glass transition. The spin glass state for higher value of $x$ is observed by others as well [9]. Magnetic field dependent magnetization, $M$ ($H$), for all the samples measured at 2 K is shown in Fig. S2 in supplemental material [16], whereas an enlarged view of the $M$ ($H$) curves is shown in Fig. 2b. For clarity, all curves are shifted along the $H$ axis. It can be seen that a clear tiny hysteresis loop is developed from a linear behavior with increasing $x$ from $x = 0.05$ to 0.14. The shape of the tiny hysteresis loop (shown in supplemental material) as observed in $x = 0.14$ is a typical feature of a spin glass system, while the linear behavior in $x = 0.05$ indicates antiferromagnetic ordering.

To investigate the nature of the magnetic interactions, $1/\chi_{\text{molar}}$ vs. $T$ data, measured under 1 kOe magnetic field, fitted with the Curie-Weiss law is shown in Fig. S3 in supplemental material [16]. The parameters obtained from the fitting are shown in Table II in the supplemental material. The effective paramagnetic moment ($\mu_{\text{eff}}$) remains ~ 4.6 $\mu_B/Co^{2+}$ for all the samples, irrespective of the amount of disorder, indicating that the cobalt ions in the octahedral sites are present in +2 oxidation state. The calculated effective paramagnetic moment for a free $Co^{2+}$ ion is 3.87 $\mu_B$ assuming a spin-only contribution and it is 6.5 $\mu_B/Co^{2+}$ when it includes orbital magnetic moments. The experimental observation of an intermediate value of $\mu_{\text{eff}}$ suggests the presence of orbital contribution. The obtained Curie - Weiss temperature ($\theta_{CW}$) for all the samples is close to -95 K and does not vary much with different amount of disorder. The negative sign of $\theta_{CW}$ indicates that the interaction is antiferromagnetic in nature. The value of $\theta_{CW}$ is very high compared to $T^*$ and the frustration parameter ($f = |\theta_C|/T^*$) is ~ 10. This indicates that the system is highly frustrated because of the competing nearest-neighbor ($J_1$) and next-nearest-neighbor ($J_2$) interactions.

Temperature dependent heat capacity ($C_p$) for $x$ = 0.05 and 0.14 are shown in Fig. 2c. A peak is observed around $T^*$ ~ 9 K for $x$ = 0.05, indicating the presence of a possible long - range magnetic ordering of $Co^{2+}$ spins. However, the nature of the peak does not resemble a conventional $\lambda$-type, as expected for a second order magnetic phase transition. It is worth mentioning here that the frustrated spinel $FeSc_2S_4$ exhibits a similar broad peak in heat capacity around 8 K and it has been suggested recently that it remains in the near but in the long - range ordered side of the critical point, which separates the spin - orbital ordering and spin - orbital liquid state [19]. Upon increasing $x$ from $x$ = 0.05 to 0.14, this peak shifts to lower temperature (~ 7.5 K) and becomes broad. This is consistent with the fact that for a spin glass system, temperature dependent heat capacity exhibits a broad peak at ~ 1.4 $T_{SG}$, where the $T_{SG}$ is the spin glass transition temperature [20]. For further analysis, phonon contribution, $C_{Phonon}$, to the total heat capacity is calculated by fitting $C_p$ ($T$) data from 50 K to 100 K using a combined Debye - Einstein model, as given below [21-23],

$$C_{Phonon} = C_{Debye} + C_{Einstein} = \frac{9Ra_1}{x_D^3}\int_0^{x_D} \frac{x^4 e^x}{(e^x - 1)^2} dx + 3R\sum_{n=1}^{2} b_n \frac{x_{E,n}^2 e^{x_{E,n}}}{(e^{x_{E,n}} - 1)^2}$$

where $R$ is the universal gas constant, $x_{D,E}$ = $\Theta_{D,E}/T$ where $\Theta_{D,E}$ is the Debye and Einstein temperature respectively. The data can be fitted well with one Debye term and two Einstein term. In the combined Debye - Einstein model, the total number of modes is same as the number of atoms in the formula unit. For $CoAl_2O_4$, the used values of Debye and Einstein coefficients $a_1$, $b_1$ and $b_2$ are 1, 1 and 5, respectively [21,22]. The obtained phonon contribution to $C_p$ ($T$) is represented by the solid lines in Fig. 2c. The phonon contribution is then subtracted from the total heat capacity and the obtained magnetic contribution to the heat capacity, $C_{\mathrm{Mag}}$ is shown in Fig. 2d as $C_{\mathrm{Mag}}/T$ vs. $T$ plot. It is clear that the magnetic contribution to the heat capacity decreases with increasing $x$ and the shift of the peak is clearly visible in Fig. 2d. Change in entropy associated with this magnetic phase transition is calculated using the equation given by,

$$\Delta S_{\mathrm{Mag}}(T) = \int_0^T \frac{C_{\mathrm{Mag}}(T)}{T} dT$$

Temperature dependent change in magnetic entropy, $\Delta S_{\mathrm{Mag}}$ ($T$), is shown in the inset of the Fig. 2d. Expected change in entropy due to magnetic ordering of $Co^{2+}$ ions with $S$ = 3/2, is $\Delta S_{Mag}$ (at high $T) = R\, ln(2S + 1)$ = 11.5 J/mole-K. For $x$ = 0.05, $\Delta S_{Mag}(at\ T = 100\ K)$ ~ 8.4 J/mole-K, which is almost 73% of the expected total entropy. With increasing $x$ from 0.05 to 0.14, $\Delta S_{\mathrm{Mag}}$ decreases to 7.9 J/mole-K, as expected for appearance of spin glass state. We note that even in $Co_3O_4$, where a complete long - range collinear antiferromagnetic ordering occurs around 30 K, the observed magnetic entropy change is ~ 70% of the expected value [21].

Temperature dependent ac susceptibility ($\chi'$) data for all the samples are shown in Fig. 3. As can be seen in Fig. 3a, a broad frequency independent peak is observed around 14 K for $x$ = 0.05 and a sharp peak is found around $T^*$ ~ 9 K in $d\chi'/dT$ (not shown), similar to the dc magnetization. A weak kink is observed around 4 K which is also frequency independent, as shown in the inset of Fig. 3a. With increasing $x$, the broad peak shifts to lower temperature and the weak kink becomes prominent. For $x$ = 0.07, a small frequency dependent shift is observed around the kink at 4 K. In $x$ = 0.14, the broad peak vanishes completely and the weak kink becomes a sharp peak around 5 K, which depends strongly on the frequency of the applied ac magnetic field, confirming the presence of a spin glass state below 5 K [20].

Thus, dc magnetization, heat capacity and ac susceptibility data indicates that the $x$ = 0.05 sample possesses a magnetic transition around $T^*$ ~ 9 K. However, it is not possible to conclude the exact nature of this magnetic transition using these measurements, whether it is long - range collinear antiferromagnetic ordering or spiral spin - liquid state. On lowering temperature below 4 K, this sample might have a spin glass state, as indicated by the weak kink in ac susceptibility data, where the frequency dependent shift could be very small. On the other hand, in the case of $x$ = 0.14, a clear spin glass transition is observed around 5 K without long - range magnetic ordering.

Temperature dependent dielectric constant $\varepsilon_r$ ($T$) measured for all the samples at 100 kHz with different magnetic fields are shown in Fig. 4. For $x$ = 0.05, no dielectric anomaly is observed around $T^*$ in the absence of magnetic field. When external magnetic field is applied, a sharp $\lambda$-shaped peak is observed around $T^*$ and its magnitude increases with increasing magnetic field, indicating the presence of a strong magnetodielectric effect. A corresponding peak in dielectric loss is also observed only in the presence of magnetic field, which remains as low as 0.02 throughout the measurement temperature range (not shown). As seen from Fig. 4, the magnetic field induced dielectric peak is observed for all $x$, however there is a difference in the nature of the peaks. With increasing disorder ($x$), the peak becomes broad and the $\lambda$-shaped nature of the peak vanishes. Another important observation is that with increasing $x$, the peak position also decreases to lower temperature, which is in accordance with the shift of the broad peak in dc and ac magnetization, as shown in Fig. 2 and 3. Also, the magnetodielectric effect decreases with increasing $x$. This becomes clear when magnetic field dependent dielectric constant is measured at fixed temperatures. In Fig. S4 (supplemental material), magnetic field dependent magnetocapacitances, MC (%), measured at the temperatures where the magnetic field induced peak in $\varepsilon_r$ ($T$) is observed, are shown [16]. In $x$ = 0.05, the observed MC (%) is ~ 2.8 % while it decreases to ~ 0.08 % for $x$ = 0.14.

Pyroelectric current measurements have been carried out to find out whether or not this magnetic field induced dielectric peak is associated with the magnetoelectric effect. Temperature dependent pyroelectric current in the presence of different magnetic fields are shown in the insets of Fig. 5 (a-c) for $x$ = 0.05, 0.07 and 0.11, respectively. The pyroelectric current does not exhibit any peak when it is measured in the absence of magnetic field for all the samples. However, a sharp peak is observed in the presence of applied magnetic fields for $x$ = 0.05, 0.07 and 0.11, similar to that observed in dielectric constants. Electric polarization is calculated with integrating pyroelectric current data with measurement time and is shown in Fig. 5, where the electric polarization is found to increase with increasing magnetic field. However, similar to the behavior of magnetic field induced dielectric peak, the natures of the pyroelectric current peak are different for different samples. For $x$ = 0.05, the pyroelectric current peak is very sharp and asymmetric. For $x$ = 0.07 and 0.11, the peaks become broad and almost symmetric. In $x$ = 0.14, the magnetic field does not induce pyroelectric current peak at all.

In Fig. 5d, we have shown the magnetic field dependence of electric polarization at 3 K, as obtained from Fig. 5 (a-c). It can be seen that the electric polarization varies linearly with magnetic field for $x \leq 0.11$. The magnetoelectric coefficient calculated from the slope of the linear fit of the $P$ vs. $H$ data is found to be 2.7 ps/m for $x$ = 0.05, which is comparable to that of $Co_3O_4$, but higher than $MnGa_2O_4$, [15]. With increasing $x$, the value of magnetoelectric coefficient decreases and vanishes for $x$ = 0.14.

In Fig. 6, we have shown the results of dc-biased current measurements on all the samples with a magnetic field of 80 kOe and an electric field of $E_{\text{Bias}}$ = + 6.5 kV/cm. It has been suggested recently that dc-biased current measurements can distinguish the pyroelectric current peak associated with ferroelectric transition from thermally stimulated free charge (TSFC) current in multiferroics [24]. It can be seen that for $x$ = 0.05, dc-biased current exhibits an upward and a downward peak, indicating the magnetoelectric polarization and depolarization, respectively. With increasing $x$, these peaks become broad and shift to lower temperature and finally vanishes for $x$ = 0.14. These results are consistent with the pyroelectric current measurements and further confirm the magnetoelectric effect.

The observation of magnetic field induced dielectric anomaly and electric polarization confirm that the samples, $Co_{1-x}Al_x[Al_{2-x}Co_x]O_4$ ($x$ = 0.05, 0.07 and 0.11) are linear magnetoelectrics. To the best of our knowledge, till today there is no material reported to be magnetoelectric without long - range magnetic ordering. Thus, the observation of linear magnetoelectric effect in $CoAl_2O_4$ indicates that the broad peaks in dc and ac magnetization are associated with a collinear antiferromagnetic ordering. In our recent work [15], using magnetoelectric measurements on single crystals of *A*-site magnetic spinel ($MnGa_2O_4$) and phenomenological calculations, we have shown that among the two possible magnetic space groups $I4'_1/a'm'd$ and $R$-$3'm'$, which are associated with the direction of the $Mn^{2+}$ spins along [100] and [111], respectively, $R$-$3'm'$ is the exact magnetic space group. However, both of these two magnetic space groups allow linear magnetoelectric effect. Based on the second order nature of the magnetic phase transition, we suggest that the present system should have any one of these two magnetic space groups. A further and accurate neutron diffraction experiments or a similar magnetoelectric measurement on single crystal of $CoAl_2O_4$ are necessary for further confirmation.

In order to support these findings, we performed Monte Carlo (MC) simulations of magnetic ordering and magnetoelectric effects in $CoAl_2O_4$ with various degrees of inversion using the Heisenberg Hamiltonian

$$H = J_1 \sum_{\langle ij \rangle} \boldsymbol{S}_i \cdot \boldsymbol{S}_j + J_2 \sum_{\langle\langle ij \rangle\rangle} \boldsymbol{S}_i \cdot \boldsymbol{S}_j + K \sum_i \left( S_{xi}^4 + S_{yi}^4 + S_{zi}^4 \right) - \sum_i \boldsymbol{H} \cdot \boldsymbol{S}_i,$$

where summations $\langle ij \rangle$ and $\langle\langle ij \rangle\rangle$ run over the first- and second-neighbors, respectively, $K$ is the single-ion anisotropy, and $\boldsymbol{H}$ is magnetic field. The spins $\boldsymbol{S}_i$ are classical vectors of unit length. The effect of inversion was modeled as voids appearing in the magnetic diamond lattice. We did not take into account possible $Co^{2+}$ spins at the $B$ - sites, however their necessity is discussed in supplemental material [16]. The ratio $J_2/J_1 = 0.11$ determined from neutron diffraction was used for simulations, while the exchange constant $J_1 = 3.14$ meV close to experimentally determined value 2.1 meV yielded better agreement for the temperature of magnetic phase transition [25]. The single-ion anisotropy was $K = 0.01 J_1$ [2]. The Monte Carlo studies were performed using the Metropolis algorithm and a system with dimensions of 20×20×20 cubic unit cells ($N = 64000$ Co atoms) and periodic boundary conditions. At each temperature the system was allowed to relax for $10^3$ Monte Carlo steps per spin (MCs), after which the statistics was collected also during $10^3$ MCs.

As noted above the ground state of the diamond lattice with $J_2/J_1 < 1/8$ is a collinear antiferromagnetic ordering, which can be described by the order parameter $\left(L_x, L_y, L_z\right)$ and which allows linear magnetoelectric effect due to the interaction

$$L_x\left(M_y P_z + M_z P_y\right) + L_y\left(M_z P_x + M_x P_z\right) + L_z\left(M_x P_y + M_y P_x\right), \quad (1)$$

where $(M_x, M_y, M_z)$ and $(P_x, P_y, P_z)$ are magnetization and electric polarization, respectively [15]. Taking into account that $\vec{L} = (\vec{S}_{A1} - \vec{S}_{A2})/2$ and $\vec{M} = (\vec{S}_{A1} + \vec{S}_{A2})/2$, where $\vec{S}_{A1}$ and $\vec{S}_{A2}$ are spins of the two Co atoms in the primitive unit cell of $CoAl_2O_4$, it follows from Eq. (1) that the ME effect is due to single-ion contribution

$$P_x \sim S_{A1y}S_{A1z} - S_{A2y}S_{A2z},$$

$$P_y \sim S_{A1x}S_{A1z} - S_{A2x}S_{A2z}, \qquad (2)$$

$$P_z \sim S_{A1x}S_{A1y} - S_{A2x}S_{A2y}.$$

Equations (2) are used in the following to calculate local spin-induced electric dipole moments.

Fig. 7 shows results of the Monte Carlo simulations. As follows from Fig. 7(a-d), the CAO:0.05 is almost completely ordered at low temperatures (the value $L = 1$ corresponds to complete antiferromagnetic ordering). With increasing inversion the AF order decreases and disappears between 11 and 14% inversion. Fig. 7(e-h) show temperature dependence of the Edwards-Anderson order parameter [26,27]

$$Q(t) = \frac{1}{N}\sum_{i=1}^{N}\left|\frac{1}{t}\int_0^t \vec{S}_i(t')dt'\right|^2,$$

where integration is replaced by summation over MCs in simulations. It follows that for CAO:0.05 the appearance of AF order is accompanied by the growth of Q(1000). The samples with higher inversion (e.g. 14%) do not show AF order upon decreasing temperature, nevertheless Q(1000) increases similarly to low inversion systems suggesting the freezing of the spins. Fig. 7(i-l) show electric polarization in systems with different degrees of inversion. Upon decreasing temperature electric polarization appears in presence of magnetic field when AF order sets in. The value of electric polarization is proportional to magnetic field as shown in Fig. S5 in the supplemental material [16]. It should be noted that (i) the results of MC simulations, as shown in Fig. 7, indicate that the low magnetic field enhances AF order, and (ii) contrary to what is observed in experiments, in MC simulations dilution of magnetic lattice naturally results in gradual shift of the Curie-Weiss temperature towards zero, which is discussed in supplemental material [16].

In order to check the influence of inversion on magnetic ordering, we calculated the correlation functions $\langle \vec{S}_i \vec{S}_j \rangle$ as functions of the relative distance $\left|\vec{R}_i - \vec{R}_j\right|$, which are shown in Fig. 8. In accordance with the results for the long - range AF order parameter $L$ the system with 5% inversion shows saturation of spin-spin correlation function at non-zero level for high spin separations. The form of the correlation function corresponds to AF order. With increasing inversion, the correlation length gradually decreases and for 14% inversion, it reaches a value that corresponds to the lattice constant meaning that the spins are not correlated beyond that distance and there is no long - range order.

Thus, our experimental results supported by MC calculations show the presence of linear magnetoelectric effect in $CoAl_2O_4$ with low inversion. $CoAl_2O_4$ lies at the border between the AF and spin spiral ground states [1], which is the reason of contradicting results obtained by various groups. The existence of linear magnetoelectric effect in spinels with the clear AF state was experimentally found earlier [15], whereas the possibility of electric polarization induced by the spiral spin states was theoretically suggested in Ref. [28]. Indeed, electric polarization in incommensurate magnetically ordered states is ubiquitous in magnetoelectrics, however the presence of linear magnetoelectric effect in the spiral spin state, which is possibly suggested for $CoAl_2O_4$, is not possible due to the following. Assuming that $J_2/J_1$ is slightly greater than 1/8, the spiral wave vector is close to the point **q**=0 [1]. This allows describing the spiral by spatially varying $\vec{L}$ and $\vec{M}$. In the macroscopic continuous description, the linear magnetoelectric effect would require the presence of macroscopic interactions of the form

$$\frac{\partial L_i}{\partial x_j} H_k P_l,$$

or

$$\frac{\partial M_i}{\partial x_j} H_k P_l,$$

where $\vec{H}$ is magnetic field. However, the first of them is forbidden because it changes sign upon inversion, whereas both of them would average to zero upon integration over volume. Therefore, the presence of linear magnetoelectricity in $CoAl_2O_4$ is the evidence of the long - range AF ordering.

## IV. Conclusion

We have shown that the compound, $Co_{1-x}Al_x[Al_{2-x}Co_x]O_4$ with low anti-site disorder ($x \leq 0.11$) exhibits linear magnetoelectric effect below the magnetic ordering temperature. The sample with $x$ = 0.14 exhibits a spin glass behavior without magnetoelectric effect. The presence of linear magnetoelectric effect, supported by Monte Carlo simulations and theoretical considerations, suggest that the ground state of $CoAl_2O_4$ with low site disorder is a long range collinear antiferromagnetic ordering. Thus, the compound $CoAl_2O_4$ with low cation disorder can be placed on the $J_2/J_1 \lesssim 1/8$ side of the phase diagram.

## Acknowledgements

A. S. would like to acknowledge Science and Engineering Board (SERB Sanction No. EMR/2014/000896), Department of Science & Technology, Government of India, for financial support. S. G. acknowledges Jawaharlal Nehru Centre for Advanced Scientific Research for providing a research fellowship. A. S. and S. G. thank Sheikh Saqr Laboratory at Jawaharlal Nehru Centre for Advanced Scientific Research for providing experimental facilities. N. V. T. acknowledges support by the Ministry of Education and Science of the Russian Federation (state assignment grant No. 3.5710.2017/BCh).

## Figure Captions:

Figure 1 (a) Crystal structure of a normal cubic spinel, $A^{2+}[B_2^{3+}]X_4$ with space group $Fd$-3$m$ (b) Diamond lattice formed by the $A$ - site magnetic ions. Green lines represent nearest-neighbor ($J_1$) and red lines represent next-nearest-neighbor ($J_2$) exchange interactions.

Figure 2 (a) Temperature-dependent field-cooled (FC) and zero-field-cooled (ZFC) magnetization of all the samples measured with a magnetic field of 100 Oe. FC curves are represented by closed symbols and ZFC curves are shown with open symbols. Inset shows temperature dependence of the first derivative ($dM/dT$) of FC magnetization with temperature for $x$ = 0.05 and 0.07. (b) Magnetic field dependent magnetization of all samples measured at 2 K. All curves are shifted along $H$ - axis for clarity. (c) Temperature-dependent heat capacity ($C_p$) measured with $H$ = 0 Oe for $x$ = 0.05 and 0.14. Solid lines represent phonon contributions to the total heat capacity obtained using a Debye-Einstein model. (d) Temperature dependent magnetic contribution to the total heat capacity ($C_{Mag}/T$) for $x$ = 0.05 and 0.14. Inset shows the entropy change associated with the magnetic transitions for $x$ = 0.05 and 0.14.

Figure 3 (a-d) Temperature dependent ac susceptibility ($\chi'$) for $x$ = 0.05, 0.07, 0.11 and 0.14, respectively. The insets show the magnified view of the low temperature peak.

Figure 4 (a-d) Temperature dependent dielectric constant measured with 100 kHz in the presence of different magnetic fields for $x$ = 0.05, 0.07, 0.11 and 0.14, respectively.

Figure 5 (a-c) Temperature dependent electric polarizations obtained from pyroelectric current measurements for $x$ = 0.05, 0.07 and 0.11, respectively. The insets show the corresponding temperature dependent pyroelectric current data, which is measured in the presence of different magnetic fields after a magnetoelectric poling with an electric field of $E_P$ = + 6.5 kV/cm and same magnetic field. (d) Magnetic field dependent electric polarization for $x$ = 0.05, 0.07 and 0.11, obtained from Fig. 5 (a-c). The magnetoelectric coefficients obtained from the linear fitting of $P$ vs. $H$ curves are mentioned in the figure.

Figure 6 Temperature dependent dc-biased current of all the samples measured with an electric field of $E_{\mathrm{Bias}}$ = + 6.5 kV/cm and a magnetic field of 80 kOe.

Figure 7 Temperature dependence of the antiferromagnetic order parameter $L$ (a-d), Edwards-Anderson order parameter $Q(1000)$ (e-h), and electric polarization (i-l) as obtained by MC simulations for all samples.

Figure 8 Spin-spin correlation functions as functions of relative distance measured in units of the cubic lattice constant for all samples.

Figure 1

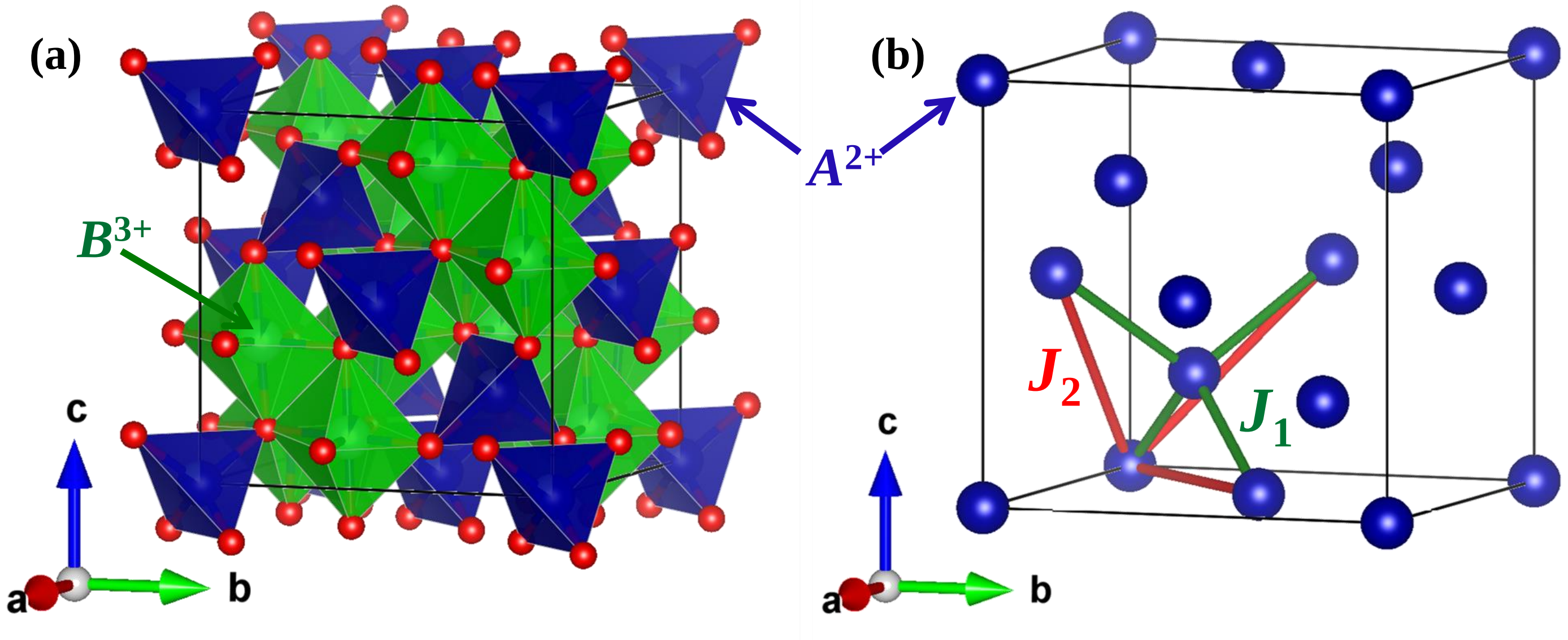

Figure 2

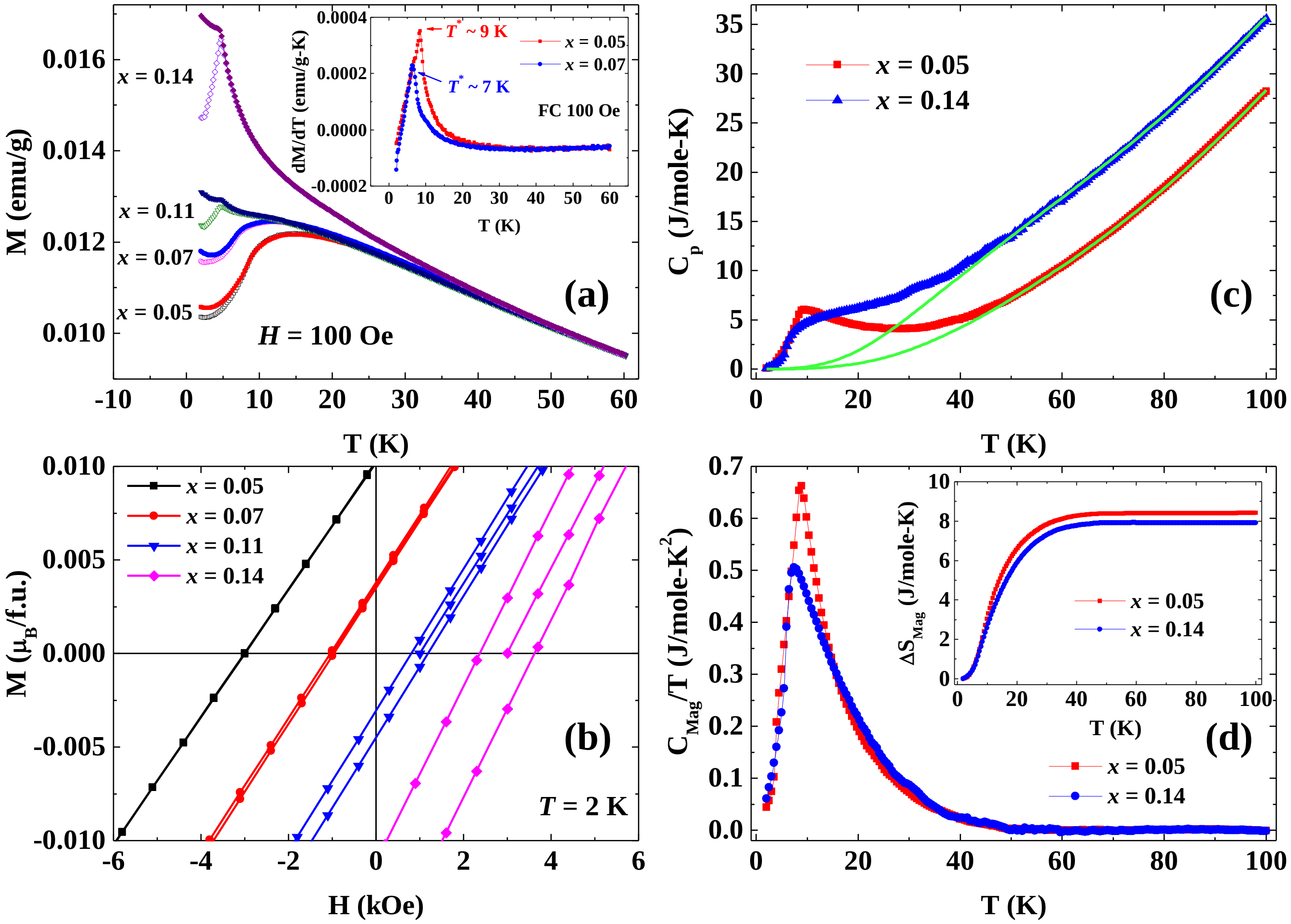

Figure 3

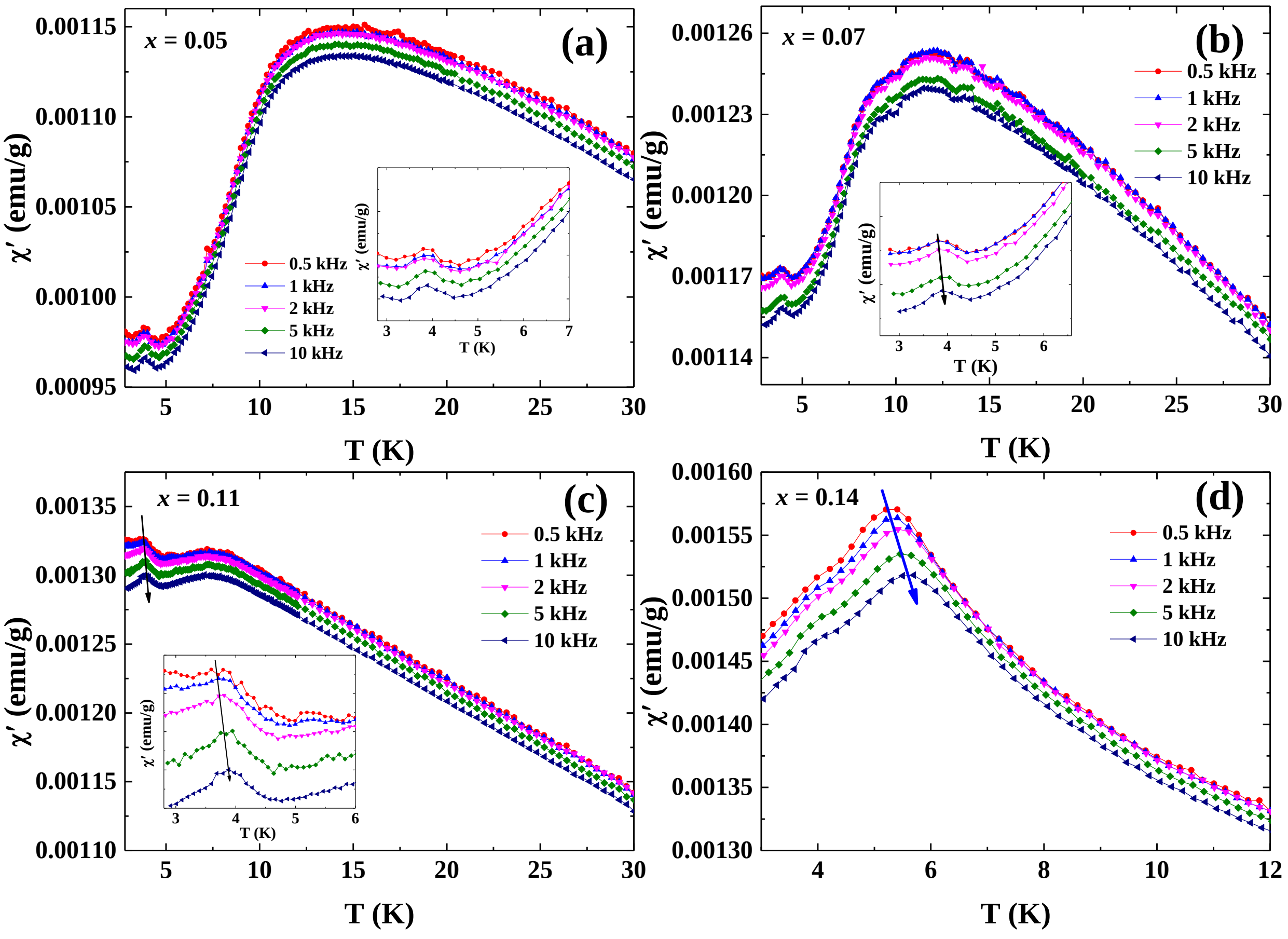

Figure 4

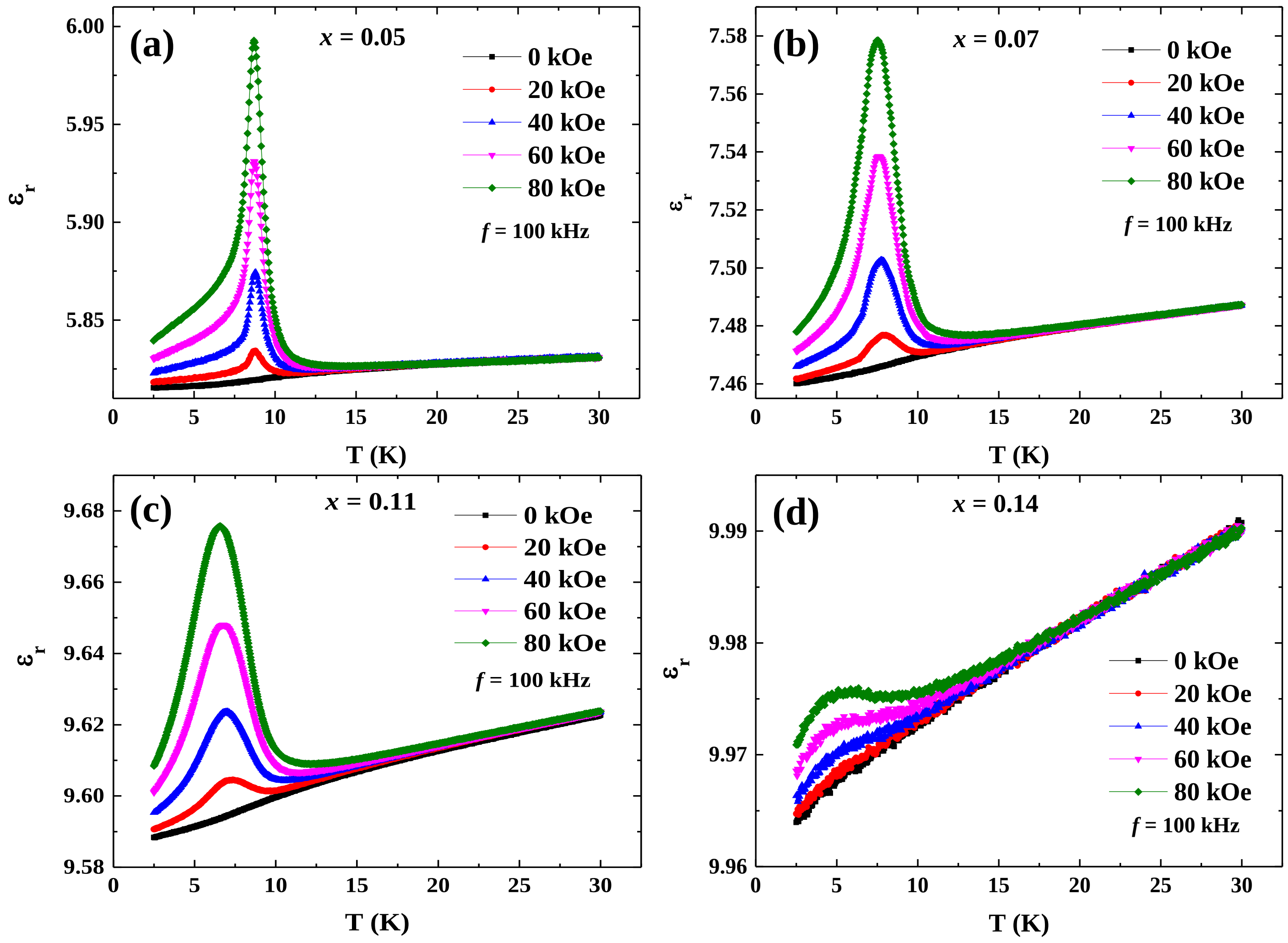

Figure 5

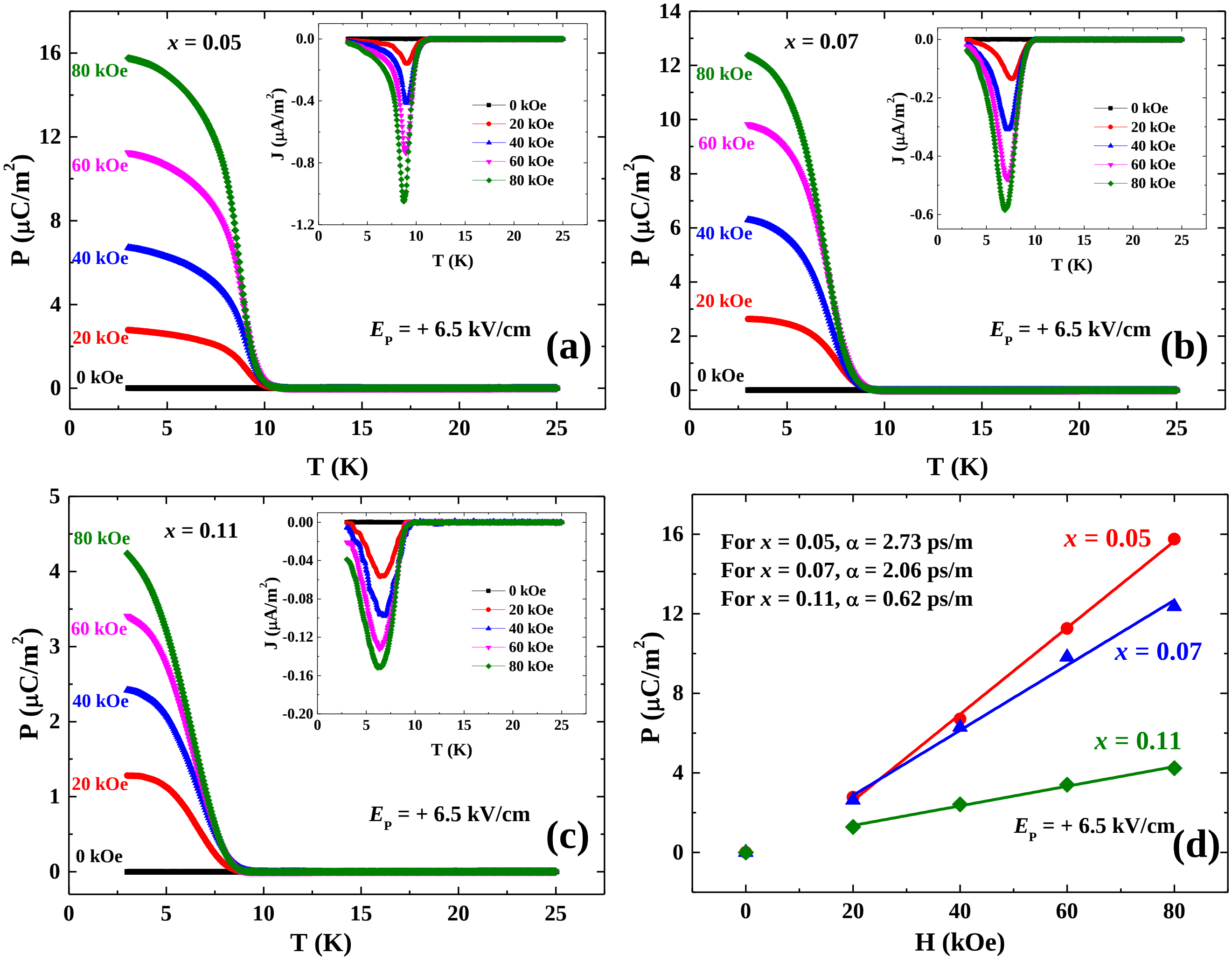

x = 0.05
80 kOe
60 kOe
40 kOe
20 kOe
0 kOe
E_P = + 6.5 kV/cm
(a)
P (μC/m²)
T (K)
J (μA/m²)
0 kOe
20 kOe
40 kOe
60 kOe
80 kOe
x = 0.07
(b)
x = 0.11
(c)
For x = 0.05, α = 2.73 ps/m
For x = 0.07, α = 2.06 ps/m
For x = 0.11, α = 0.62 ps/m
x = 0.05
x = 0.07
x = 0.11
E_P = + 6.5 kV/cm
(d)
H (kOe)

Figure 6

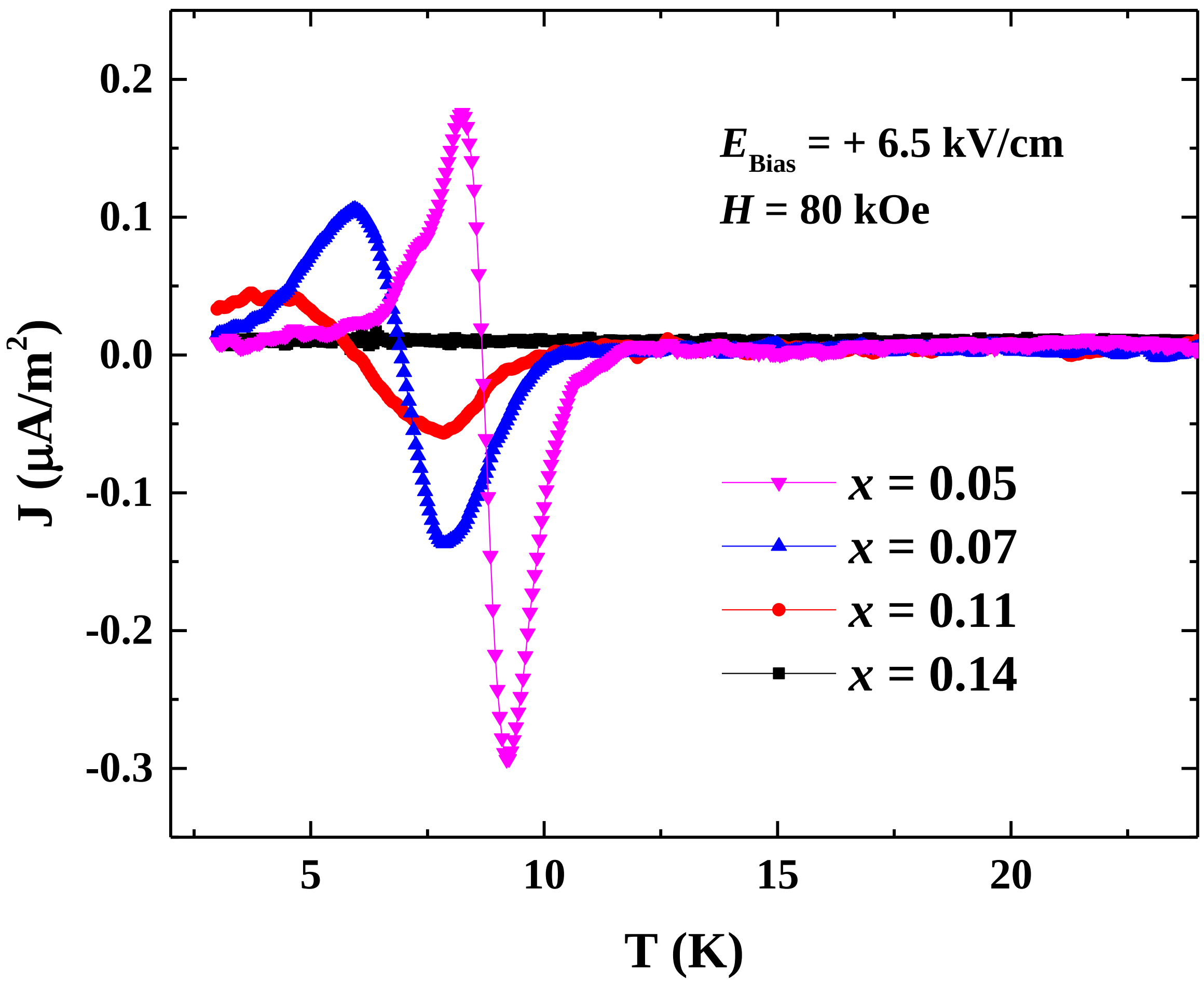

Figure 7

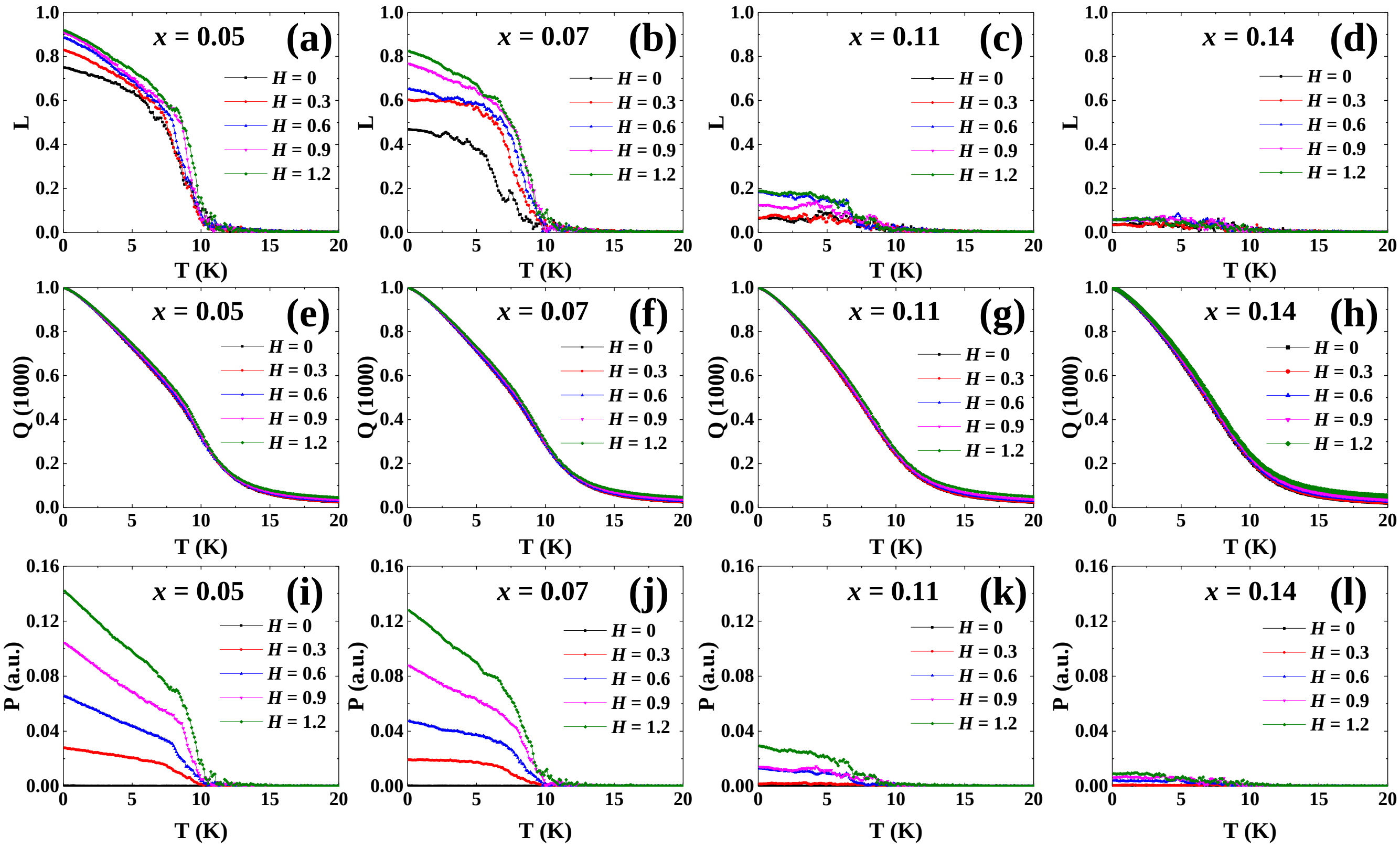

Figure 8

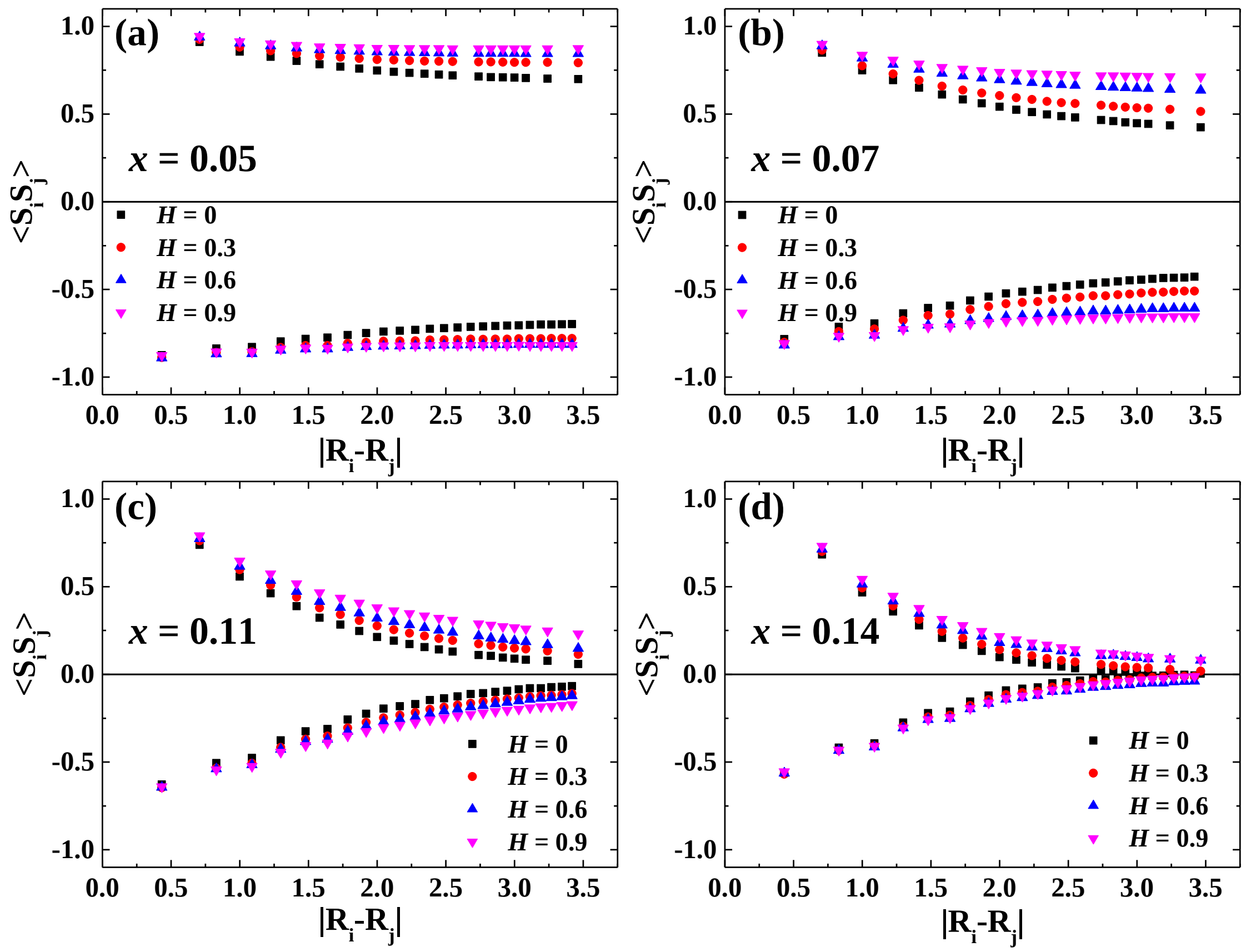

(a)
x = 0.05
H = 0
H = 0.3
H = 0.6
H = 0.9
(b)
x = 0.07
H = 0
H = 0.3
H = 0.6
H = 0.9
(c)
x = 0.11
H = 0
H = 0.3
H = 0.6
H = 0.9
(d)
x = 0.14
H = 0
H = 0.3
H = 0.6
H = 0.9
<SiSj>
|Ri-Rj|

# Supplemental Material

# Linear magnetoelectric effect as a signature of long-range collinear antiferromagnetic ordering in the frustrated spinel $CoAl_2O_4$

Somnath Ghara,[1] N. V. Ter-Oganessian,[2] and A. Sundaresan[1,*]

[1]Chemistry and Physics of Materials Unit and International Centre for Materials Science, Jawaharlal Nehru Centre for Advanced Scientific Research, Jakkur P.O., Bangalore 560 064, India.

[2]Institute of Physics, Southern Federal University, Rostov-on-Don 344090, Russia.

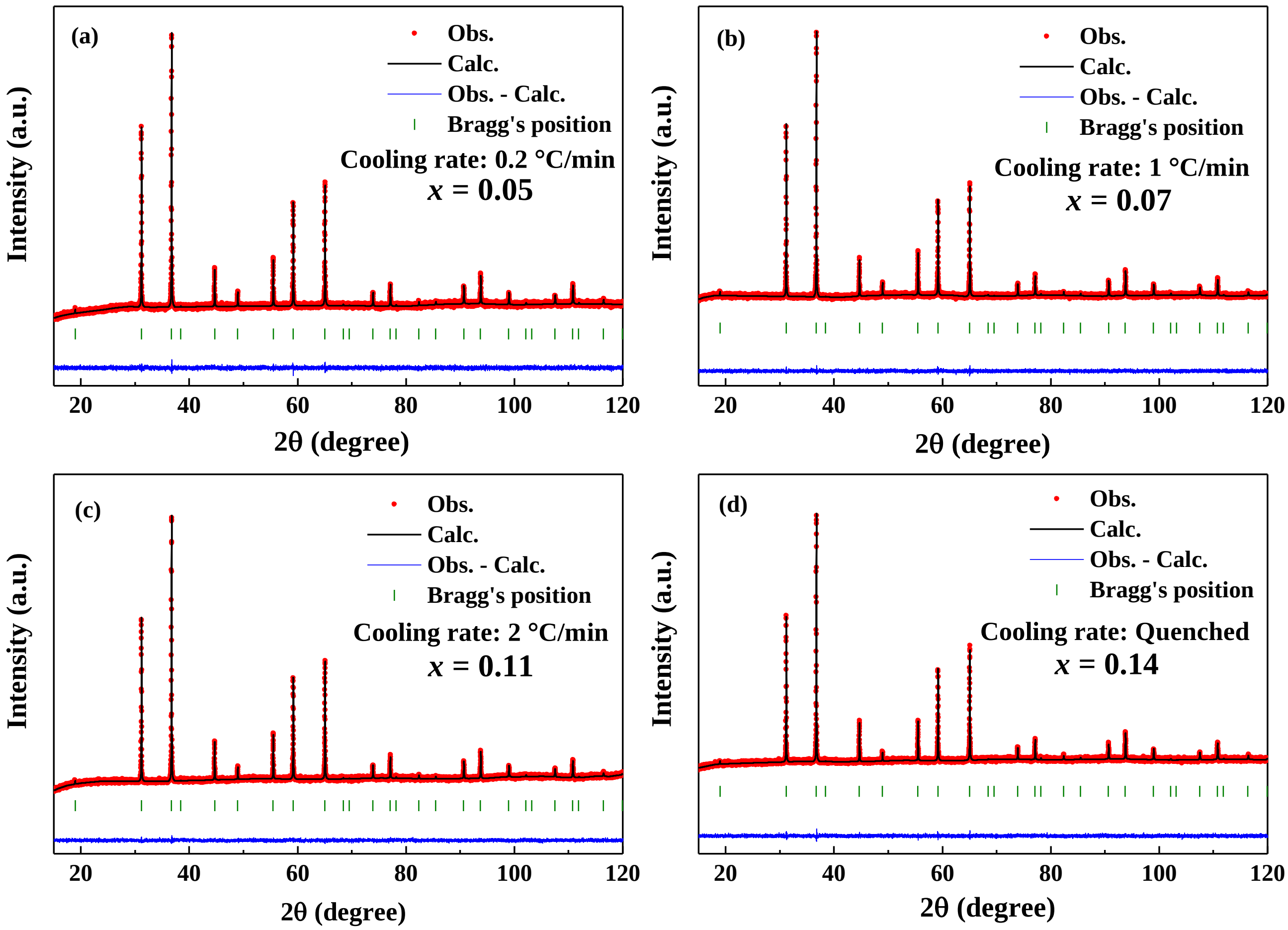

Fig. S1 Room temperature X-ray diffraction patterns of $Co_{1-x}Al_x[Al_{2-x}Co_x]O_4$ prepared with different cooling rates. Structural parameters obtained from the Rietveld refinements are shown in Table I in the supplemental material.

Table I: Structural parameters obtained from Rietveld refinements of the room temperature X-ray diffraction pattern of $Co_{1-x}Al_x[Al_{2-x}Co_x]O_4$ (Space group: *Fd*-3*m*)

$x$ = 0.05; $a = b = c$ = 8.1069 (1) Å

| Atom | $x$ | $y$ | $z$ | $B_{ios}$ (Å$^2$) | Occupancy |
|---|---|---|---|---|---|
| Co | 1/8 | 1/8 | 1/8 | 0.33 ( 7) | 0.95 (1) |
| Co | 1/2 | 1/2 | 1/2 | 0.33 ( 7) | 0.05 (1) |
| Al | 1/8 | 1/8 | 1/8 | 0.32 (11) | 0.05 (1) |
| Al | 1/2 | 1/2 | 1/2 | 0.32 (11) | 1.95 (1) |
| O | 0.2645 (3) | 0.2645 (3) | 0.2645 (3) | 0.14 (11) | 4.00 |

Bragg $R$ - factor = 4.08, $\chi^2$ = 1.10

$x$ = 0.07; $a = b = c$ = 8.1070 (1) Å

| Atom | $x$ | $y$ | $z$ | $B_{ios}$ (Å$^2$) | Occupancy |
|---|---|---|---|---|---|
| Co | 1/8 | 1/8 | 1/8 | 0.40 (7) | 0.93 (1) |
| Co | 1/2 | 1/2 | 1/2 | 0.40 (7) | 0.07 (1) |
| Al | 1/8 | 1/8 | 1/8 | 0.38 (10) | 0.07 (1) |
| Al | 1/2 | 1/2 | 1/2 | 0.38 (10) | 1.93 (1) |
| O | 0.2643 (2) | 0.2643 (2) | 0.2643 (2) | 0.27 (10) | 4.00 |

Bragg $R$ - factor = 2.93, $\chi^2$ = 1.11

$x$ = 0.11; $a = b = c$ = 8.1072 (1) Å

| Atom | $x$ | $y$ | $z$ | $B_{ios}$ (Å$^2$) | Occupancy |
|---|---|---|---|---|---|
| Co | 1/8 | 1/8 | 1/8 | 0.24 (6) | 0.89 (1) |
| Co | 1/2 | 1/2 | 1/2 | 0.24 (6) | 0.11 (1) |
| Al | 1/8 | 1/8 | 1/8 | 0.58 (11) | 0.11 (1) |
| Al | 1/2 | 1/2 | 1/2 | 0.58(11) | 1.89 (1) |
| O | 0.2643 (2) | 0.2643 (2) | 0.2643 (2) | 0.41( 9) | 4.00 |

Bragg $R$ - factor = 5.03, $\chi^2$ = 1.12

$x$ = 0.14; $a = b = c$ = 8.1075 (2) Å

| Atom | $x$ | $y$ | $z$ | $B_{ios}$ (Å$^2$) | Occupancy |
|---|---|---|---|---|---|
| Co | 1/8 | 1/8 | 1/8 | 0.18 (8) | 0.86 (1) |
| Co | 1/2 | 1/2 | 1/2 | 0.18 ( 8) | 0.14 (1) |
| Al | 1/8 | 1/8 | 1/8 | 0.19 (12) | 0.14 (1) |
| Al | 1/2 | 1/2 | 1/2 | 0.19 (12) | 1.86 (1) |
| O | 0.2634 (3) | 0.2634 (3) | 0.2634 (3) | 0.31 (11) | 4.00 |

Bragg $R$ - factor = 5.75, $\chi^2$ = 1.10

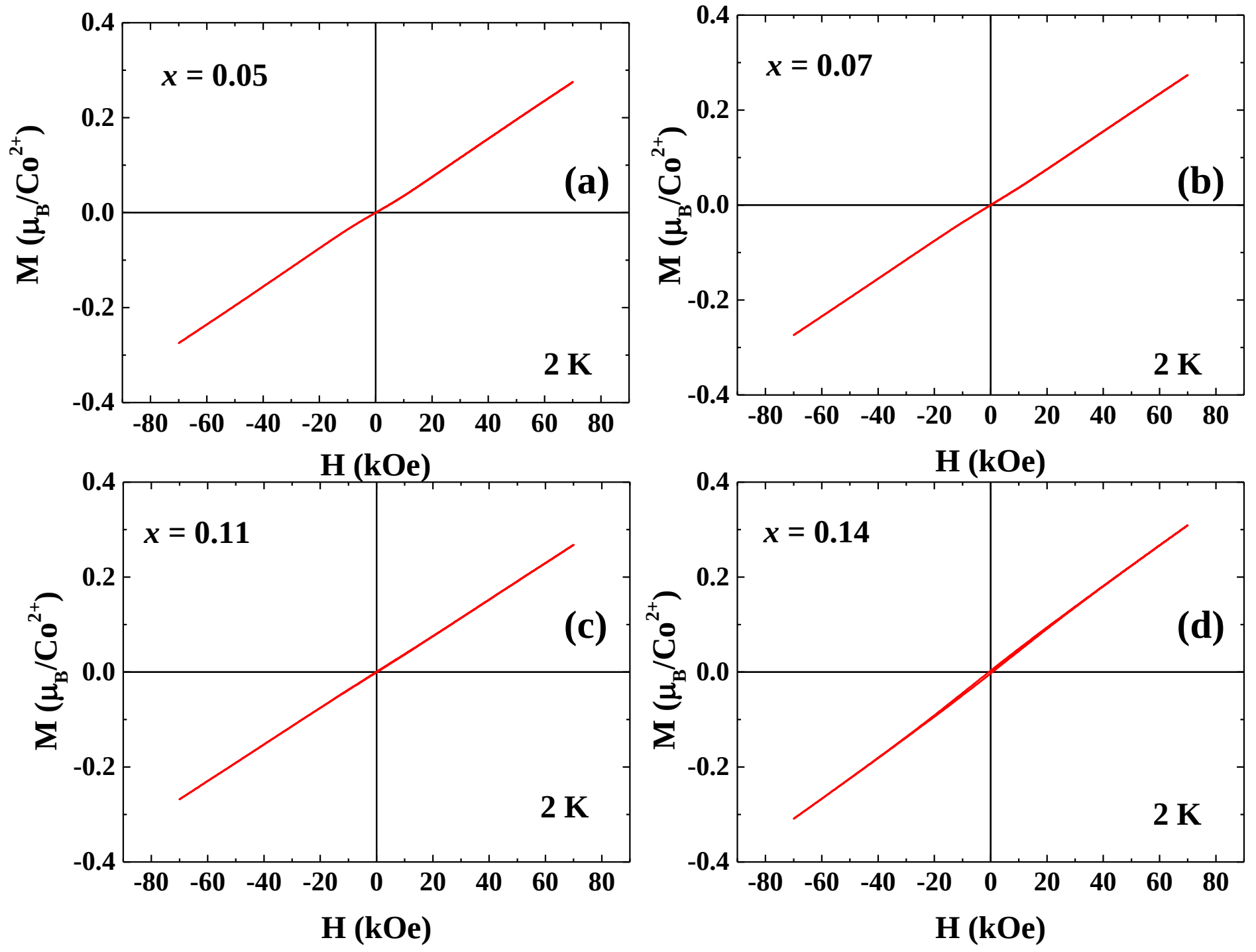

Fig. S2 Magnetic field dependent magnetization of all samples measured at 2 K

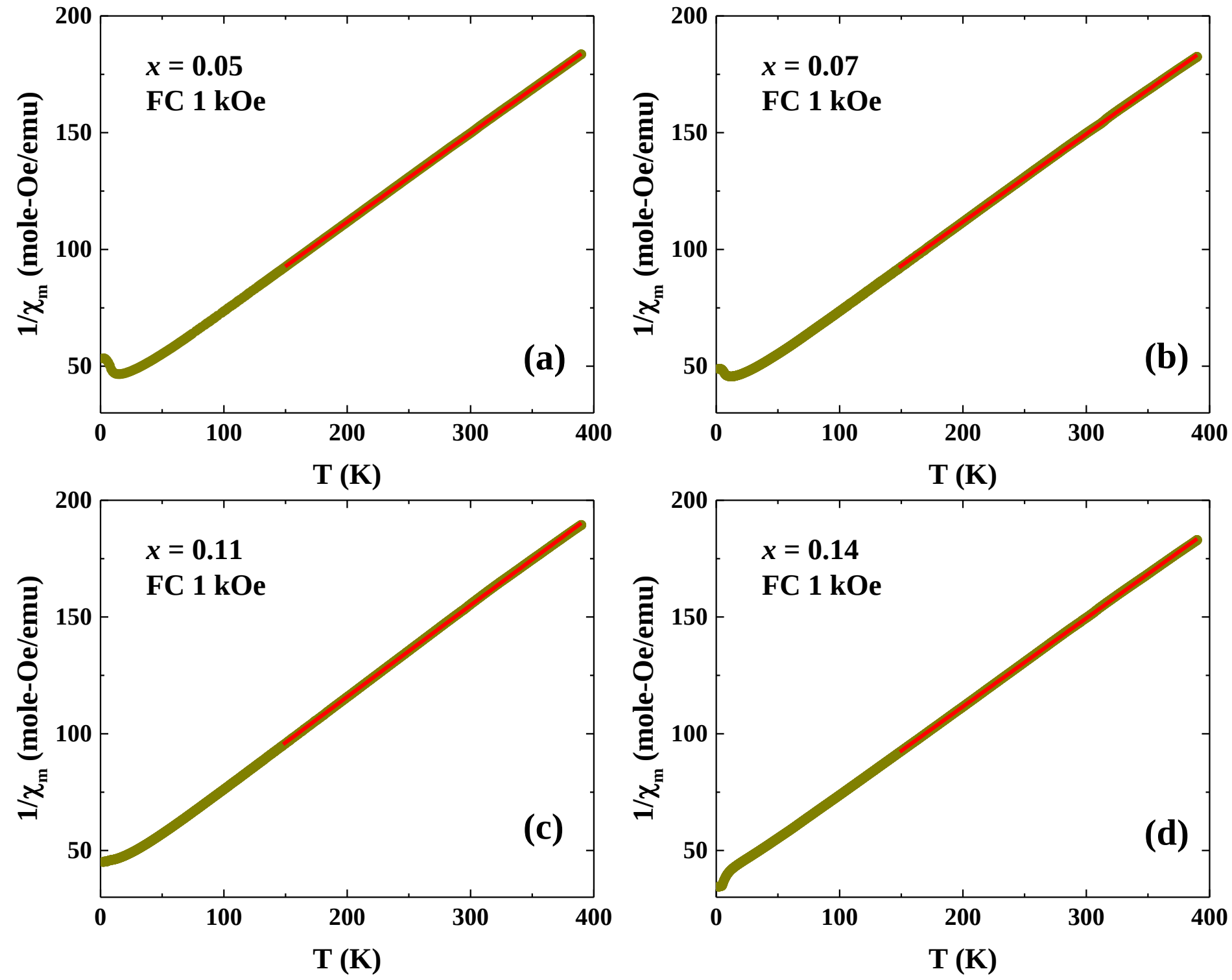

Fig. S3 Temperature dependent $1/\chi_m$ for all samples measured with a magnetic field of 1 kOe. Solid red lines show the fitting with the Curie-Weiss law.

Table II: Results of the fitting of $1/\chi_m$ vs. $T$ data from 150 K to 390 K with Curie - Weiss law for all samples of $Co_{1-x}Al_x[Al_{2-x}Co_x]O_4$

| $x$ | $\mu_{eff}$ ($\mu_B/Co^{2+}$) | $\theta_{CW}$ (K) |
|---|---|---|
| 0.05 | 4.6 | -94.4 |
| 0.07 | 4.6 | -96.9 |
| 0.11 | 4.5 | -95.4 |
| 0.14 | 4.6 | -94.8 |

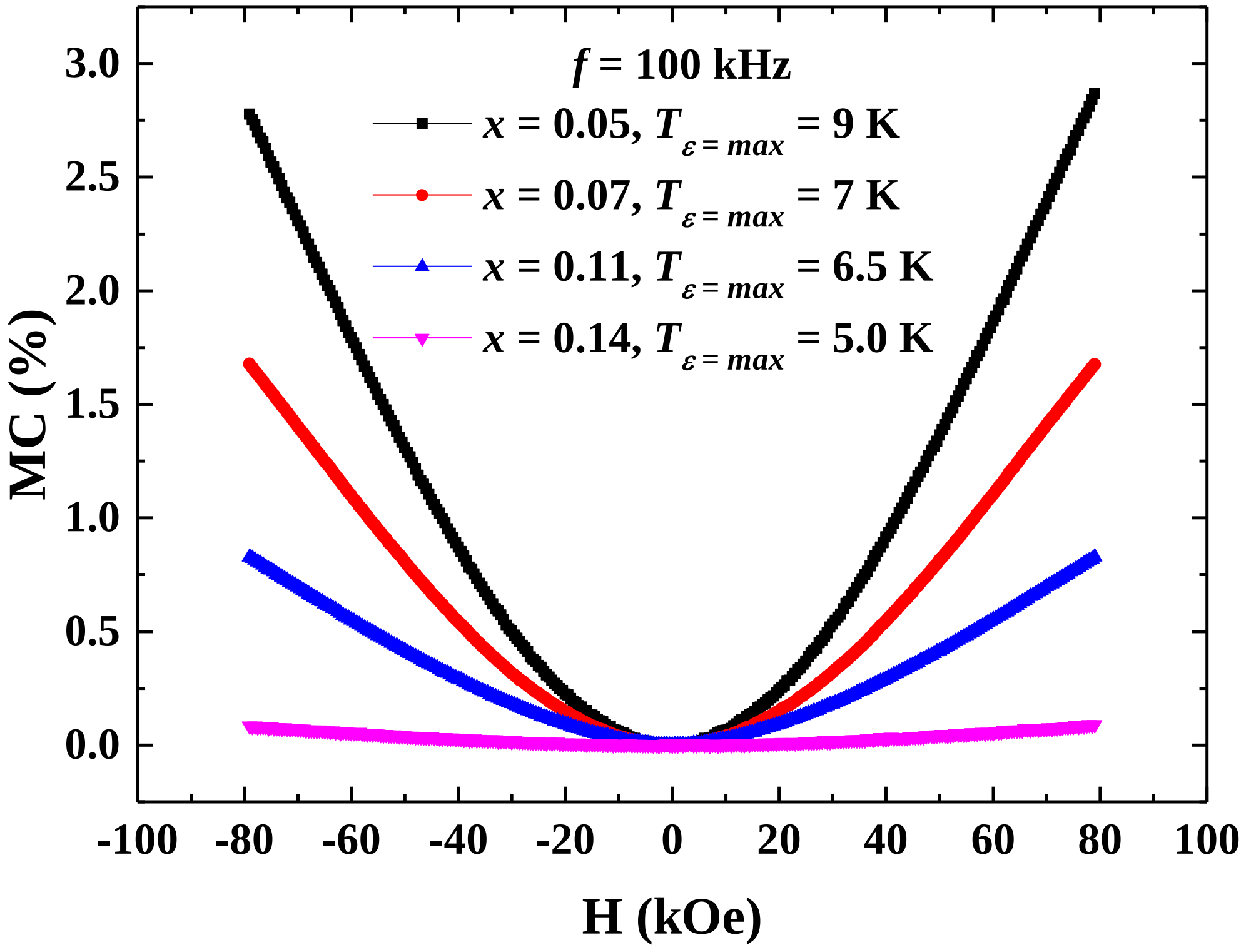

Fig. S4 Magnetic field dependent magnetocapacitance for all the samples measured with 100 kHz at the temperature, where magnetic field induced peak in $\varepsilon_r$ vs. $T$ data is observed.

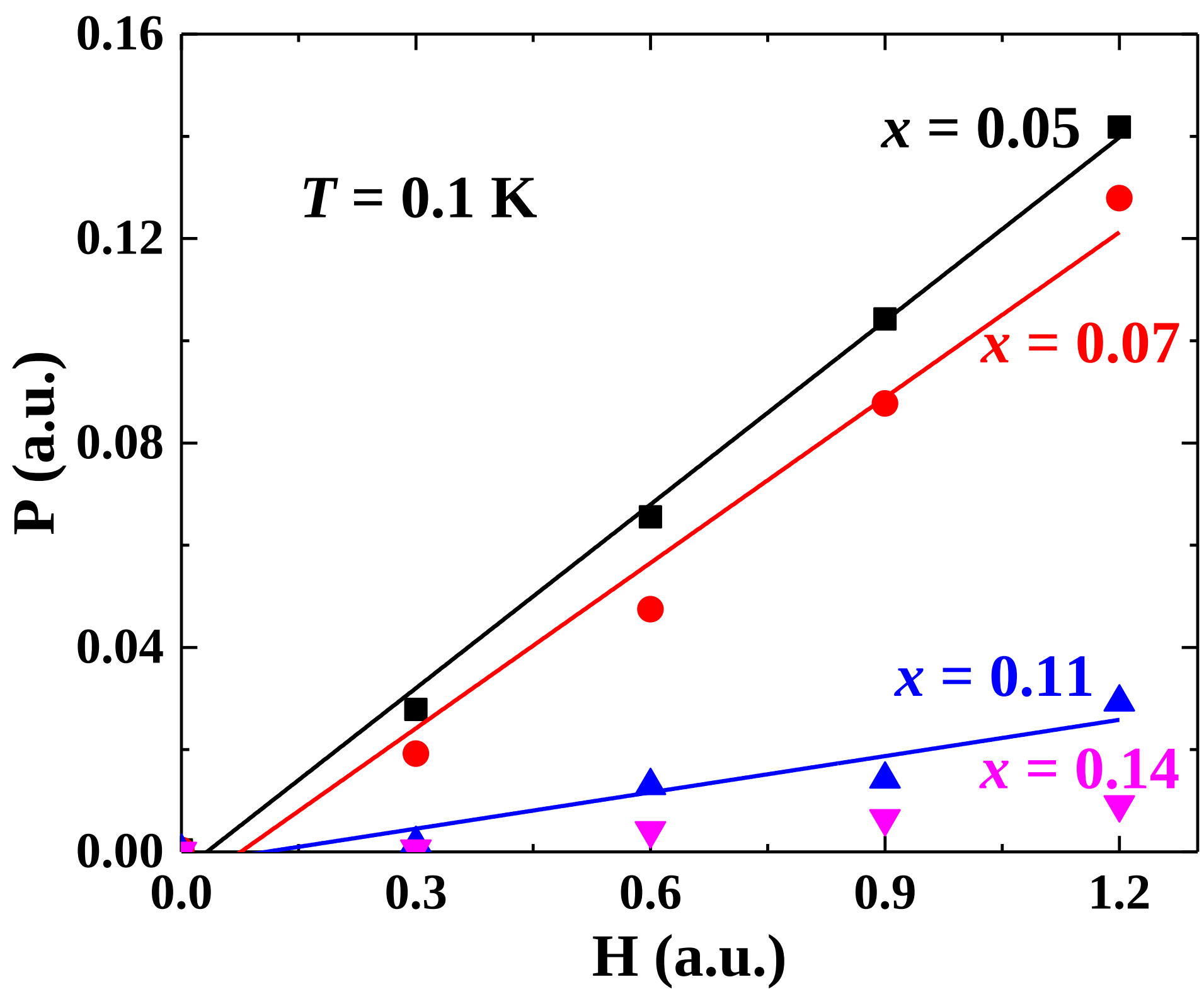

Fig. S5 Magnetic field dependent electric polarization at $T = 0.1$ K for all samples obtained from MC calculations.

## Curie-Weiss temperature of $CoAl_2O_4$

It follows from our MC calculations of high temperature magnetic susceptibility that the Curie-Weiss temperature varies linearly with inversion as shown in Fig. S6, which is in line with dilution of the magnetic lattice. Such behavior is also consistent with the experimental results on dilution of magnetic sublattice by non-magnetic ion in $Co_{1-x}Zn_xAl_2O_4$ [1]. Our experimental results as well as the results of some other studies suggest that neither effective magnetic moment nor the Curie-Weiss temperature of $CoAl_2O_4$ significantly change upon inversion up to at least $x = 0.153$ [2]. Thus, in Ref. [2] it was suggested that the Co ion in the *B*-site is also divalent with the high-spin state $S = 3/2$ $(e_g^2 t_{2g}^5)$.

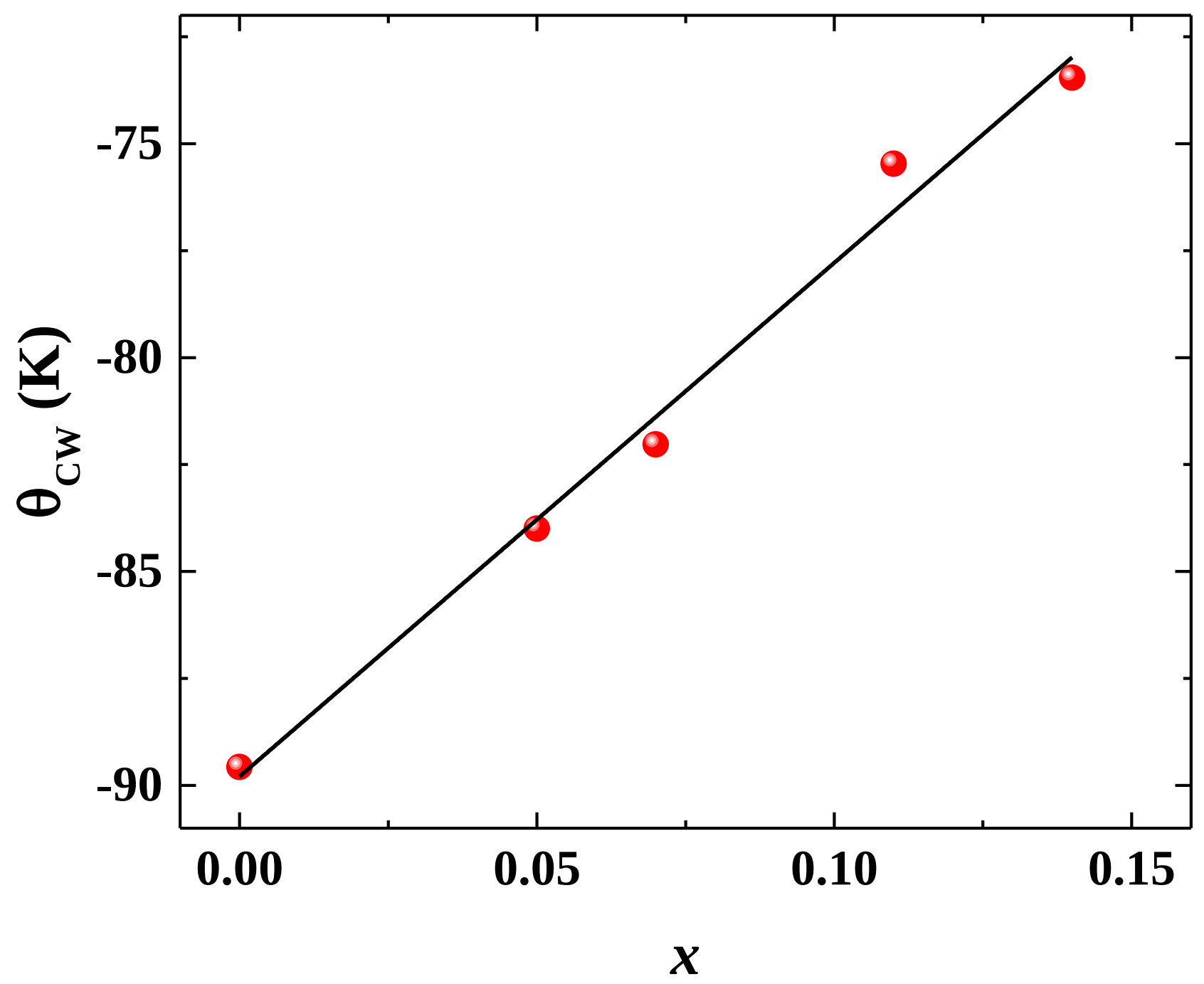

Fig. S6 Curie-Weiss temperature ($\theta_{CW}$) as function of the degree ($x$) of inversion in MC simulations. The solid line is a linear fit.

In the molecular field approximation the dependence of the Curie-Weiss temperature on inversion (for small $x$ values) can be represented as

$$\theta(x) = -\frac{4(J_1+3J_2)S(S+1)}{3k_B}\left(1 - \frac{2(J_1+3J_2)^2-9J_{AB}^2}{2(J_1+3J_2)^2}x - \frac{81J_{AB}^4}{4(J_1+3J_2)^4}x^2\right) \quad \text{(S1)}$$

where we assumed that $J_{AB}$ is the Co-Co exchange between the *A*- and *B*-sites and that the Co ions in the *A*- and *B*-sites have the same spin. Taking $J_2/J_1 = 0.11$ and $J_1 = 3.14/S^2$ meV $=$ 1.4 meV used in simulations we obtain $\theta(0) \approx -107.9$ K in fair agreement with the values obtained from both the experiments and MC simulations. From the fact that experimentally $\theta(x)$ is independent of $x$ for low inversions one can estimate the value of $J_{AB}$ requiring that the linear in $x$ term in Eq. (S1) vanishes, which gives $J_{AB}^2 = 2(J_1 + 3J_2)^2/9$. This implies that the Co-Co exchange between the *A*- and *B*-sites can be both ferromagnetic and antiferromagnetic leading to the same results. Substituting the obtained value of $J_{AB}$ into Eq. (S1) we find that $\theta(0.14)/\theta(0) \approx 0.98$ in agreement with the absence of noticeable dependence of the Curie-Weiss temperature on inversion. Thus, our findings give an estimate of $J_{AB}$, which appears comparable to $J_1$, and call for careful experimental re-examination of the exchange interactions in $CoAl_2O_4$.